\documentclass{PoS}

\title{Exclusive Photoproduction of Upsilon in pPb collisions with the CMS}

\ShortTitle{Exclusive Upsilon by CMS}

\author{\speaker{Ruchi Chudasama (for the CMS Collaboration)}\\
        Nuclear Physics Division \\
        Bhabha Atomic Research Center\\
        E-mail: \email{ruchi.physics@gmail.com}}

\author{Dipanwita Dutta\\
        Nuclear Physics Division\\
        Bhabha Atomic Research Center  \\
        E-mail: \email{dutta.dipa@gmail.com}}

\abstract{Results of exclusive photoproduction of Upsilon states in Ultraperipheral collisions (UPC) of protons and ions  with the CMS experiment are presented, which provides a clean probe of the gluon distribution at small values of parton fractional momenta $x \approx 10^{-2} 10^{4}$ at central rapidities (|y| $<  2.5$).  The three Upsilon states (1S, 2S, 3S) are measured in the dimuon decay channel along with the photon-photon decaying to dimuon QED continuum at $\sqrt{s_{NN}}=5.02$~TeV for integrated luminosity of $L_{int} = 35$ nb$^{-1}$. The total Upsilon photoproduction cross sections at different photon-proton center of mass energy $W_{\gamma p}$ and t-differential distributions are presented and compared with other experimental results as well as theoretical predictions.}

\FullConference{7th International Conference on Physics and Astrophysics of Quark Gluon Plasma\\
		1-5 February , 2015\\
		Kolkata, India}

\begin{document}

\section{Introduction}
Exclusive photoproduction of heavy vector mesons at very high photon-proton 
center-of-mass energies $(W_{\gamma p})$ can be studied in ultraperipheral collisions (UPC)
of protons (ions). Recently, CMS, ALICE~\cite{ref1} and LHCb~\cite{ref2} presented their measurements 
of exclusive heavy vector meson photoproduction at the LHC.
Since the process occurs through
$\gamma p$ or $\gamma Pb$ interaction via the exchange of two-gluons with
no net color transfer and thus, at the LO,  the cross section  is proportional
to the  square of the gluon density in the target proton or ion.
It provides a valuable probe of the gluon density at the small gluon
 momentum fraction $x$ which is kinematically related to $W_{\gamma p}$
($x=(M_{\Upsilon} /W_{\gamma p})^2$).
The  exclusive photoproduction of $\Upsilon$(1S, 2S, 3S) measurement was presented 
in their dimuon decay channel in ultraperipheral collisions 
of protons and heavy ions (pPb) with the CMS experiment at 
$\sqrt {s_{{\rm NN}}} = 5.02$ TeV for an integrated 
luminosity of $L_{{\rm int}} = 33$ nb$^{-1}$ (The luminosity corresponds to the data taken by HLT trigger).
The photoproduction cross section for $\Upsilon$$(nS)$ was measured
 as a function of $W_{\gamma p}$ in the range $91< W_{\gamma p} < 826$ GeV 
which corresponds to the rapidity of
 the $\Upsilon$ meson in the range $|y| < 2.2 $ and $x$ values are of the order $x\sim 10^{-4}$ to $x\sim 1.3\cdot 10^{-2}$.
(Later in the analysis, we decided to restrict the rapidity, |y|$<$2.2, to have better muon finding efficiency). 
 The dependence of the elastic $\Upsilon$ photoproduction cross section on the
squared $\Upsilon$ transverse momentum  
approximating the four-momentum transfer at the proton vertex ($|t|\approx p_{\rm T}^2$), 
 can be parametrized  with an exponential function $e^{-b|t|}$ at low
 values of $|t|$. The differential cross section $d\sigma/dt$, was measured in the range
 $|t| < 1.0$ (GeV/c)$^{2}$ and  the b-slope parameter was estimated.

\begin{figure*}[htbp]
\begin{center}
 \includegraphics[width=0.40\textwidth]{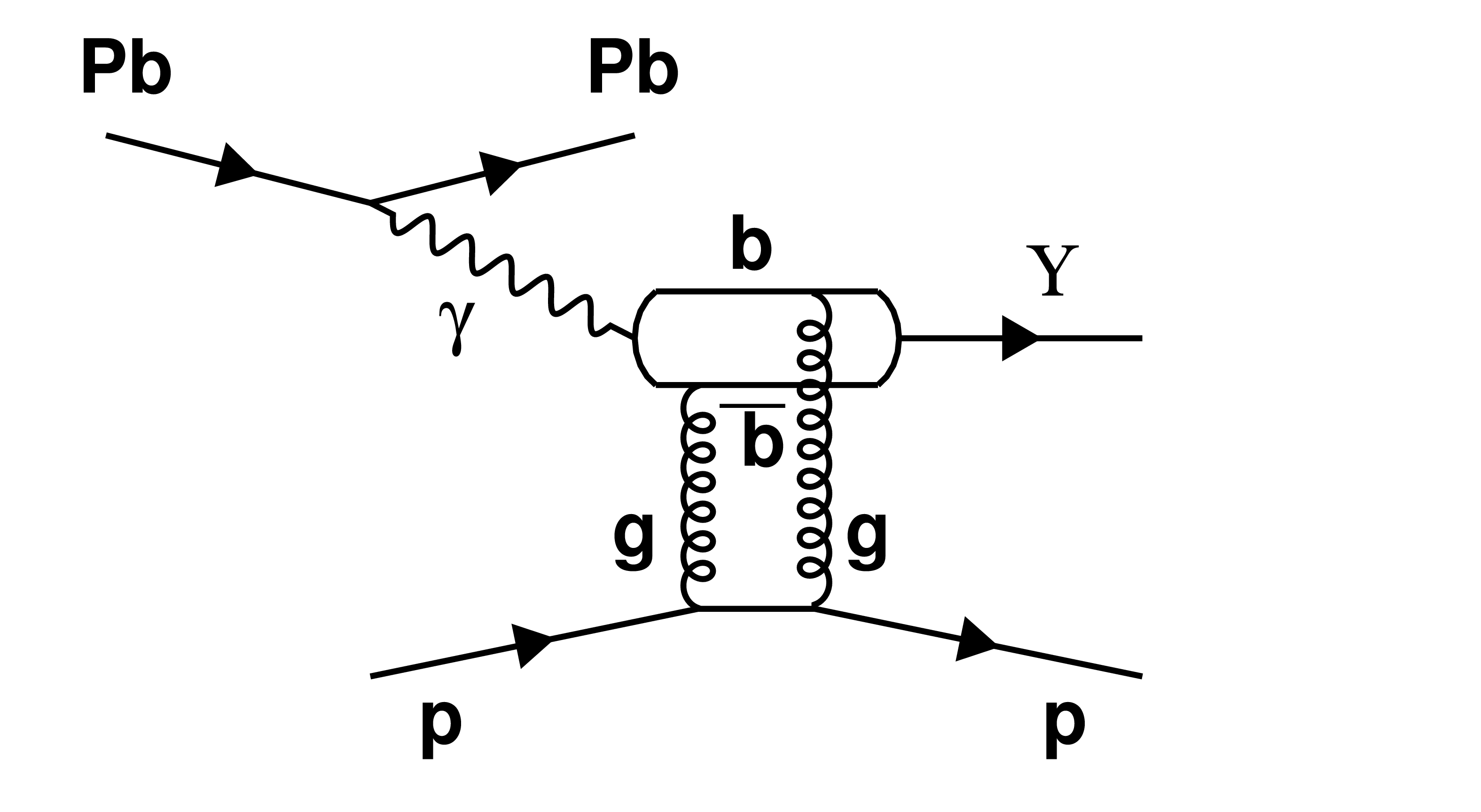}
 \includegraphics[width=0.40\textwidth]{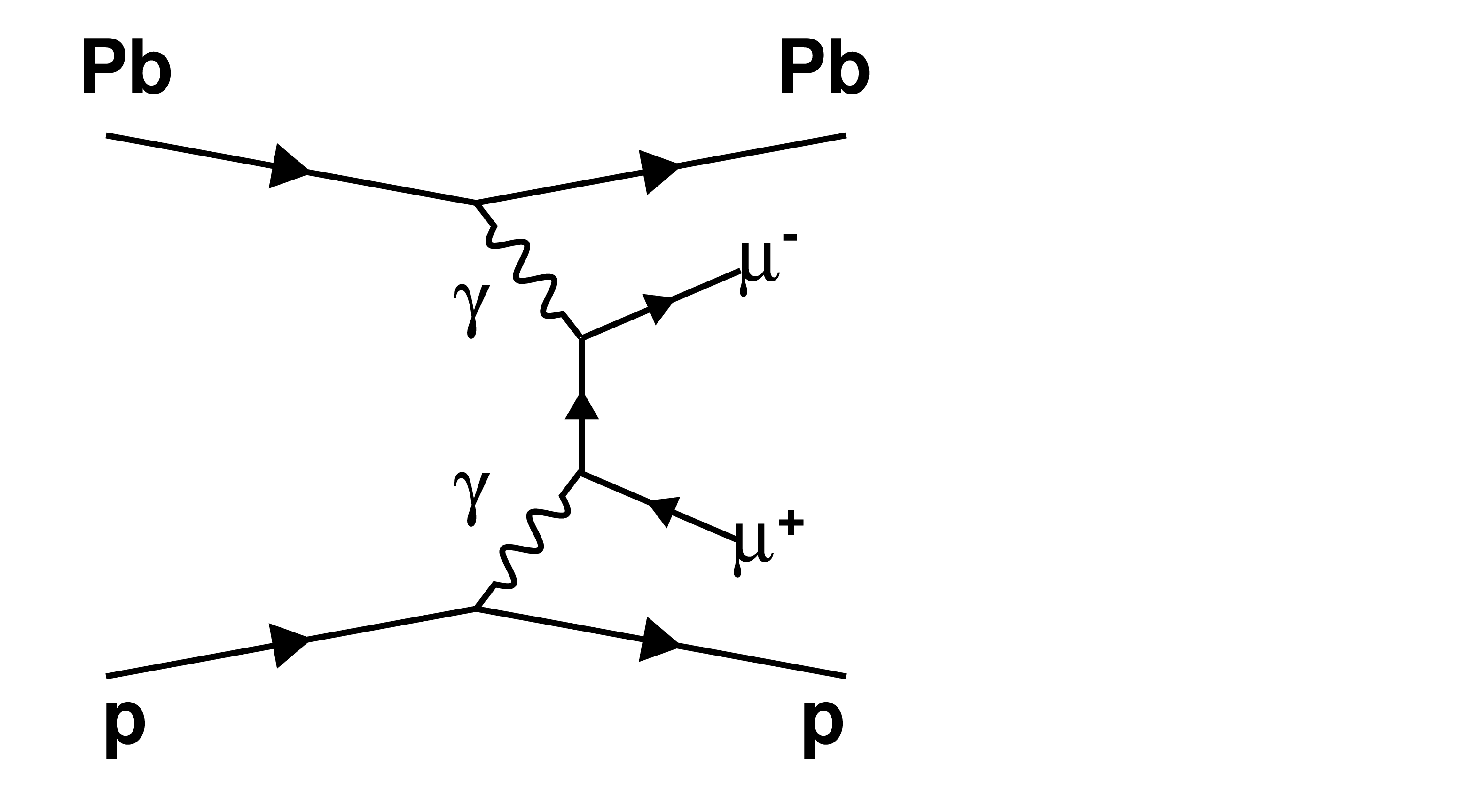}
\caption{Diagrams representing exclusive $\Upsilon$ photoproduction (left), and exclusive dimuon QED continuum (right) in pPb collisions ~\cite{ref20}.} 
\label{fig:feynman}
\end{center}
\end{figure*}

\section{Detector} 
The CMS detector at LHC is composed of superconducting solenoid of $6$m
 internal diameter which provides 
the magnetic field of $3.8$T. The solenoid volume contains silicon pixel,
strip tracker, electromagnetic calorimeter,
hadronic calorimeter. Muons are measured in gas-ionization detectors embedded 
in the steel flux-return yoke.
The muons are measured in the pseudorapidity window  $|{\eta}|< 2.5$, with muon
station consists of: Drift Tubes, Cathode Strip Chambers, and Resistive Plate Chambers.
Matching the tracks from muon stations to the tracks measured in the silicon
tracker results in a transverse
momentum resolution better than $1.5\%$ for $p_{T}$ smaller than $100$ GeV/c.
The first level (L1) of the CMS trigger system, composed of custom hardware processors, 
uses information from the calorimeters and
muon detectors to select the most interesting events in a fixed time interval of less
than $4$ $\mu$s. The high-level 
trigger (HLT) processor farm further decreases the event rate to less than 1 kHz, before data storage. A detailed description of
CMS can be found in Ref.~\cite{ref9}. 

\section{Event selection} 
The pPb collisions dataset at $\sqrt {s_{{\rm NN}}}=5.02$ recorded by CMS experiment in 2013,
corresponding to integrated luminosity $33$ nb$^{-1}$  was used in this analysis.
 The data comprise two subsets: the pPb sample,
with the Pb ion going in $+z$ direction, corresponding to an integrated
luminosity of $18.8$ nb$^{-1}$, and the Pbp sample, with the Pb ion going in the
$-z$ direction, corresponding to an integrated luminosity
of $13.8$ nb$^{-1}$. 

The exclusive $\Upsilon$(nS) photoproduction signal (Figure~\ref{fig:feynman} (left)) and
elastic QED background $\gamma \gamma \rightarrow \mu^{+}\mu^{-}$ (Figure .~\ref{fig:feynman} 
(right)) were generated using {\textsc{Starlight}} 
event generator. The exclusive signal events (where the photon comes from Pb) 
were simulated assuming an exponential dependence on the four-momentum transfer
squared at the proton vertex, $e^{-b|t|}$ with b = 4 GeV$^{2}$, 
and power-law dependence of the cross section on the photon-proton center-of-mass
 energy, $(W_{\gamma p}^{\delta})$ with $\delta = 1.7$. 
The $b$ and $\delta$ parameter were tuned to describe the data in this analysis.
The small fraction of events where the photon is emitted by proton is considered 
as background in this analysis. 

The UPC events were selected by applying dedicated HLT trigger which
selects at least one muon in each event 
and at least one to six tracks. The $\Upsilon$(nS) states are studied in
dimuon decay channel. To select the 
exclusive $\Upsilon$(nS) events, two muon tracks originating from the same
 primary vertex in each event were used.
The muons were selected with $p_{\rm T} > 3.3$ GeV and pseudorapidity $|\eta|< 2.2$, 
in order to have high muon finding efficiency.
The $p_{\rm T}$ of the muon pair was selected between $0.1$ to $1$ GeV.
The lower cut on muon pair reduces the 
contamination from elastic QED background and higher cut on muon pair 
reduces the contamination from inelastic background (proton dissociation, inclusive $\Upsilon$, Drell-Yan). The rapidity of muon pair is restricted to $2.2$. 

\begin{figure*}[hbtp]
\begin{center}
 \includegraphics[width=0.80\textwidth]{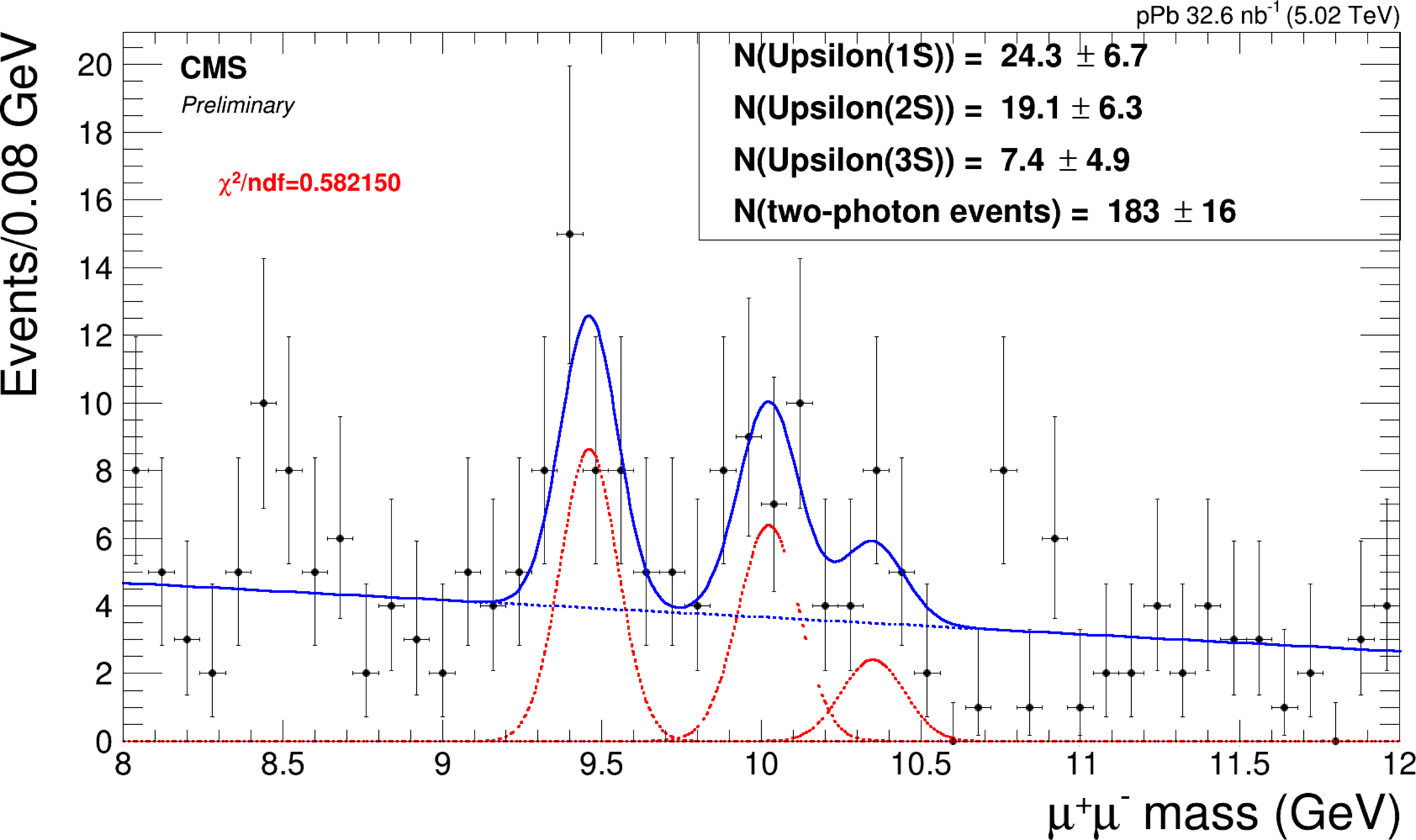}
\caption{Invariant mass distribution for the $\mu^{+}\mu^{-}$ pair in the mass range
of $8< \mu^{+}\mu^{-}< 12$ GeV/c$^2$ . The three peaks correspond to the
 $\Upsilon$(1S), $\Upsilon$(2S) and  $\Upsilon$(3S), respectively. The fit to the data is performed
with \textsc{Roofit}. The blue line corresponds to the polynomial fit to the continuum,
 the red dashed line to
the Gaussian fit to the resonances. The corresponding number of events are present in the legend~\cite{ref20}.}
\label{fig:invmass}
\end{center}
\end{figure*}

Figure~\ref{fig:invmass} shows the invariant mass distribution of the
$\mu^{+}\mu^{-}$ pair in the range between $8$ and $12$ GeV
where the three resonances $\Upsilon$(1S), $\Upsilon$(2S) and $\Upsilon$(3S), are visible. 
The fit to the data was 
performed using \textsc{RooFit}~\cite{ref7} with a linear function to describe the
continuum background from the exclusive two-photon process and three Gaussians to describe the signal peaks. The width of the $\Upsilon$(1S) 
Gaussian and the signal and background yields were free parameters,
while the widths of the $\Upsilon$(2S) and $\Upsilon$(3S) 
peaks were fixed to their world average values~\cite{ref8}. 

\begin{figure*}[hbtp]
\begin{center}
 \includegraphics[width=0.40\textwidth]{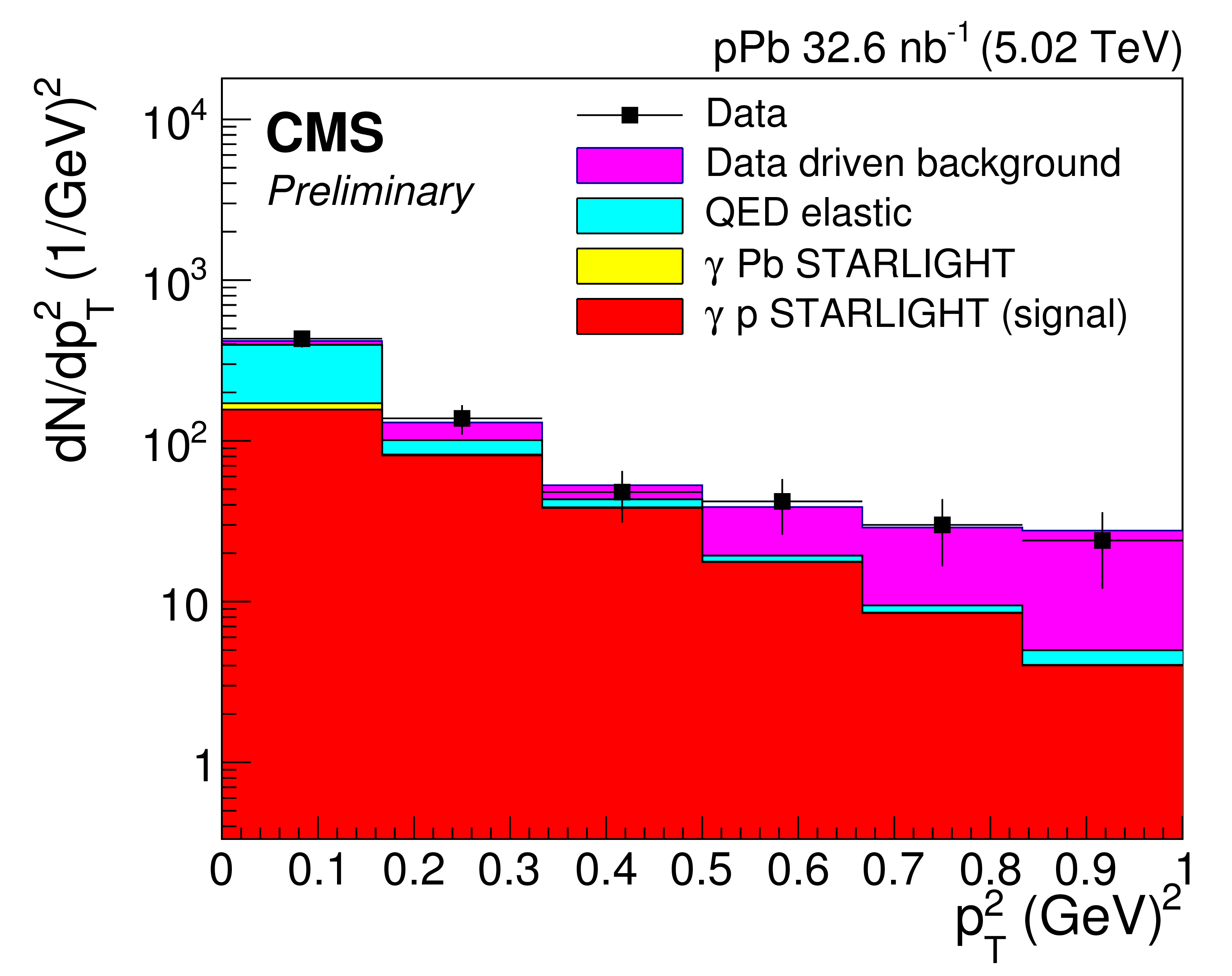}
 \includegraphics[width=0.40\textwidth]{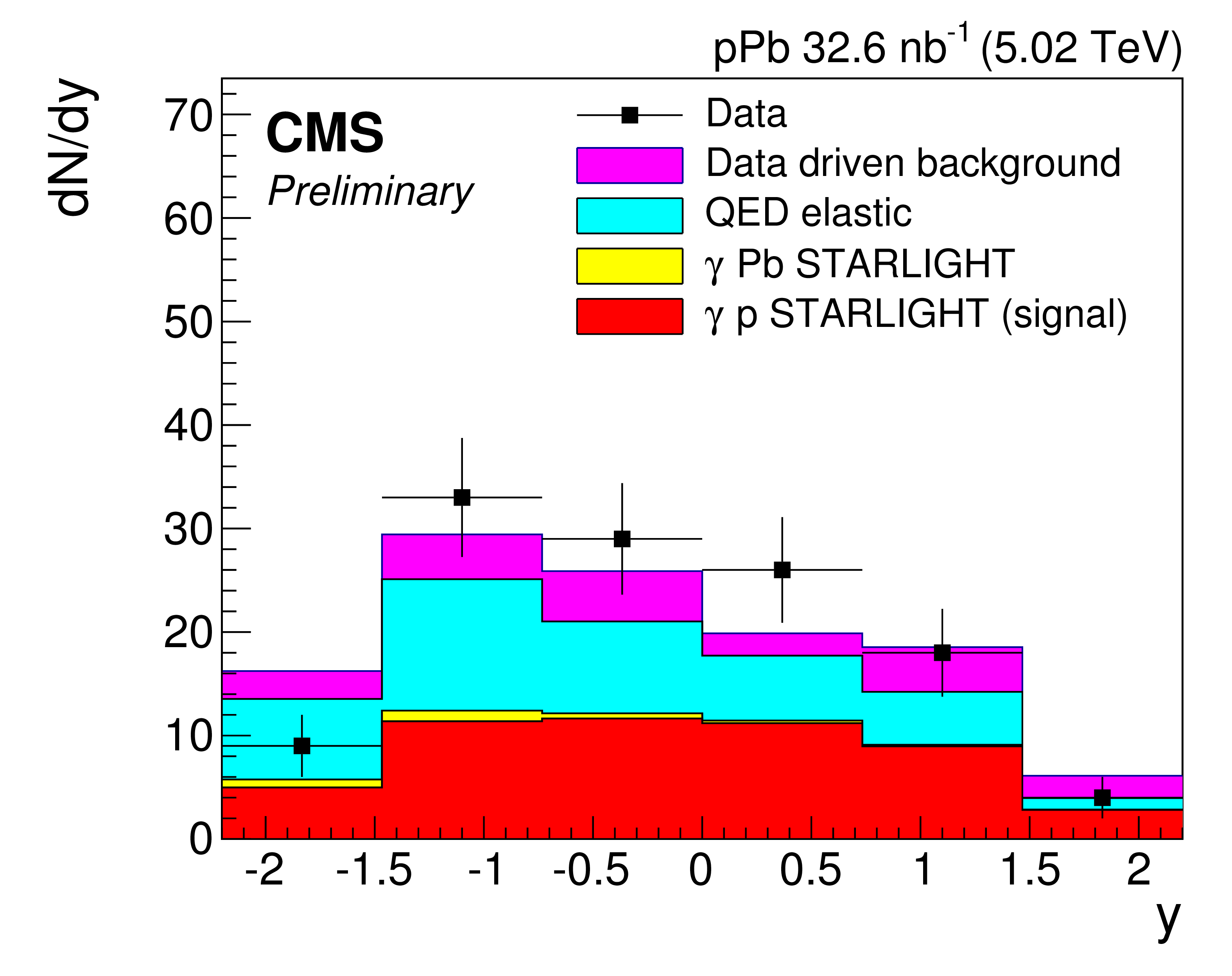}
\caption{Distributions of the transverse momentum squared $p^2_{\rm{T}}$ and rapidity $y$ of the muon pairs with
 invariant mass $9.12<m_{\mu^{+}\mu^{-}}<10.64$ GeV~\cite{ref20}.}
\label{fig:controlplots}
\end{center}
\end{figure*}

Figure ~\ref{fig:controlplots} shows the dimuon $p_{\rm T}^2$ (left) and rapidity (right) distributions in
the invariant mass interval $9.12<m_{\mu^{+}\mu^{-}}<10.64$~GeV for events passing all the selection criteria
for the combined pPb and Pbp samples. The data are compared to the MC simulations of the exclusive $\Upsilon$(nS)
production and elastic QED background processes, both simulated with \textsc{Starlight} and normalized to the
luminosity of the data. Also shown are the inelastic background events estimated from the data as explained in
Section \ref{sec:backg}. 

\section{Determining the exclusive yield}
\label{sec:backg}
The dominant background contribution to exclusive $\Upsilon$ signal comes from QED,  
$\gamma \gamma \rightarrow \mu^{+}\mu^{-}$, which was estimated by {\textsc{Starlight}}. 
The absolute prediction of QED was checked by comparing the data between invariant mass region 8--9.12 and 10.64--12 GeV  
for dimuon $p_{\rm T} < 0.15$ GeV to the simulation. The simulation reproduces data well in this region, with data-MC ratio 1.03 $\pm$ 0.10.
The contribution of non-exclusive background (inclusive $\Upsilon$, Drell-Yan and proton dissociation) was estimated by data-driven 
method by selecting events with more than 2 tracks. This template was normalized to two muon track sample in the region 
of dimuon $p_{T} > 1.5$ GeV. The normalized template describes well the region of high dimuon $p_{T}^{2}$ in the data in all four y bins used for the cross 
section extraction. Additional background in this analysis originates from a small contribution of 
exclusive $\gamma {\rm Pb} \rightarrow$ $\Upsilon$ ${\rm Pb}$ events. The fraction of these events in the total
number of exclusive $\Upsilon$ events was estimated using the reweighted \textsc{Starlight} $\Upsilon$ MC sample.
These backgrounds were subtracted from data to get the exclusive signal. 

\section{Cross section extraction}
\label{sec:xsec}
The data sample selected as described in Section 3 and 4 was used to determine the differential
${\rm d}\sigma/{\rm d}|t|$ and ${\rm d}\sigma/{\rm d}y$ cross sections in five bins of $|t|\approx p_{T}^{2}$
and four bins of $y$ respectively, for
$0.01 < |t| < 1$ GeV$^2$ and $|y| < 2.2$.
The  ${\rm d}\sigma/{\rm d}|t|$  
dsitribution was used to extract the $b$ slope of the exponential $|t|$ dependence. 
The background subtracted $|t|$ and $y$ distributions were first unfolded
to the region $0.01<|t|<1$~GeV$^2$, $|y|<2.2$, 
and muon $p_{\rm T}^{\mu}>3.3$~GeV, using the iterative Bayesian unfolding technique, as implemented in the \textsc{RooUnfold} 
package, with four iterations, to correct for detector effects  and data migration between bins. 
The ${\rm d}\sigma/{\rm d}|t|$ distribution is further extrapolated to the
full range of muon transverse momenta by means of an acceptance correction factor  
$A_{\rm corr}=N^{\Upsilon({\rm nS})}(p_{\rm T}^{\mu}>3.3$~GeV$)/N^{\Upsilon({\rm nS})}(p_{\rm T}^{\mu}>0)$, 
estimated using the \textsc{Starlight} $\gamma {\rm p}$ simulation. The cross section 
is extracted according to 
\begin{equation}
\frac{{\rm d}\sigma_{\Upsilon}}{{\rm d}|t|}=\frac{N^{\Upsilon{\rm (nS)}}}{\mathcal{L} \times \Delta |t|}~,
\end{equation}
where $|t|$ is approximated by the dimuon transverse momentum squared $p_\mathrm{T}^2$, $N^{\Upsilon{\rm (nS)}}$
denotes the background-subtracted, unfolded and acceptance-corrected number of signal events in each $|t|$ bin, 
$\mathcal{L}$ is the integrated luminosity, and $\Delta |t|$ is the width of each $|t|$ bin.

The differential $\Upsilon$(1S) photoproduction cross section ${\rm d}\sigma/{\rm d}y$ is extracted in four
bins of dimuon rapidity according to
\begin{equation}
\label{eq:cross_dsigmady}
\frac{d\sigma_{\Upsilon({\rm 1S})}}{dy}=\frac{f_{\Upsilon({\rm 1S})}}{\mathcal{B}(1+f_{\rm FD})}
\frac{N^{\Upsilon{\rm (nS)}}}{\mathcal{L} \times \Delta y}~,
\end{equation}
where $N^{\Upsilon{\rm (nS)}}$ denotes the background-subtracted, unfolded and acceptance-corrected number of
signal events in each rapidity bin. The factor $f_{\Upsilon({\rm 1S)}}$ describes the ratio of $\Upsilon${\rm
  (1S)} to $\Upsilon${\rm (nS)} events, $f_{\rm FD}$ is the feed-down contribution to the $\Upsilon${\rm (1S)}
events originating from the $\Upsilon$(2S)$\rightarrow \Upsilon({\rm 1S}) + X$ decays (where
$X=\pi^{+}\pi^{-}$ or $\pi^{0}\pi^{0}$), $\mathcal{B} = (2.48\pm 0.05)\%$ is the branching
ratio for muonic $\Upsilon(1S)$ decays, and $\Delta y$ is the width of the $y$ bin.

The $f_{\Upsilon({\rm 1S)}}$ fraction is used from the results of the inclusive $\Upsilon$ analysis~\cite{ref10}.
The feed-down contribution of $\Upsilon$(2S) decaying to $\Upsilon(1S) + \pi^{+}\pi^{-}$ and $\Upsilon(1S) + \pi^{0}\pi^{0}$ 
was estimated as $15$\%  from the \textsc{Starlight}. The contribution from feed-down of exclusive $\chi_b$ states was neglected,
as these double-pomeron processes are expected to be comparatively much suppressed in proton-nucleus collisions~\cite{ref11,ref12}.

The systematic uncertainty in the measurement of the exponential
slope $b$ of the ${\rm d}\sigma/{\rm d}|t|$  
and ${\rm d}\sigma/{\rm d}y$ distribution was considered, 
which amounts to be 13$\%$ and 25$\%$, respectively.  

\begin{figure*}[t]
\begin{center}
\includegraphics[width=0.80\textwidth]{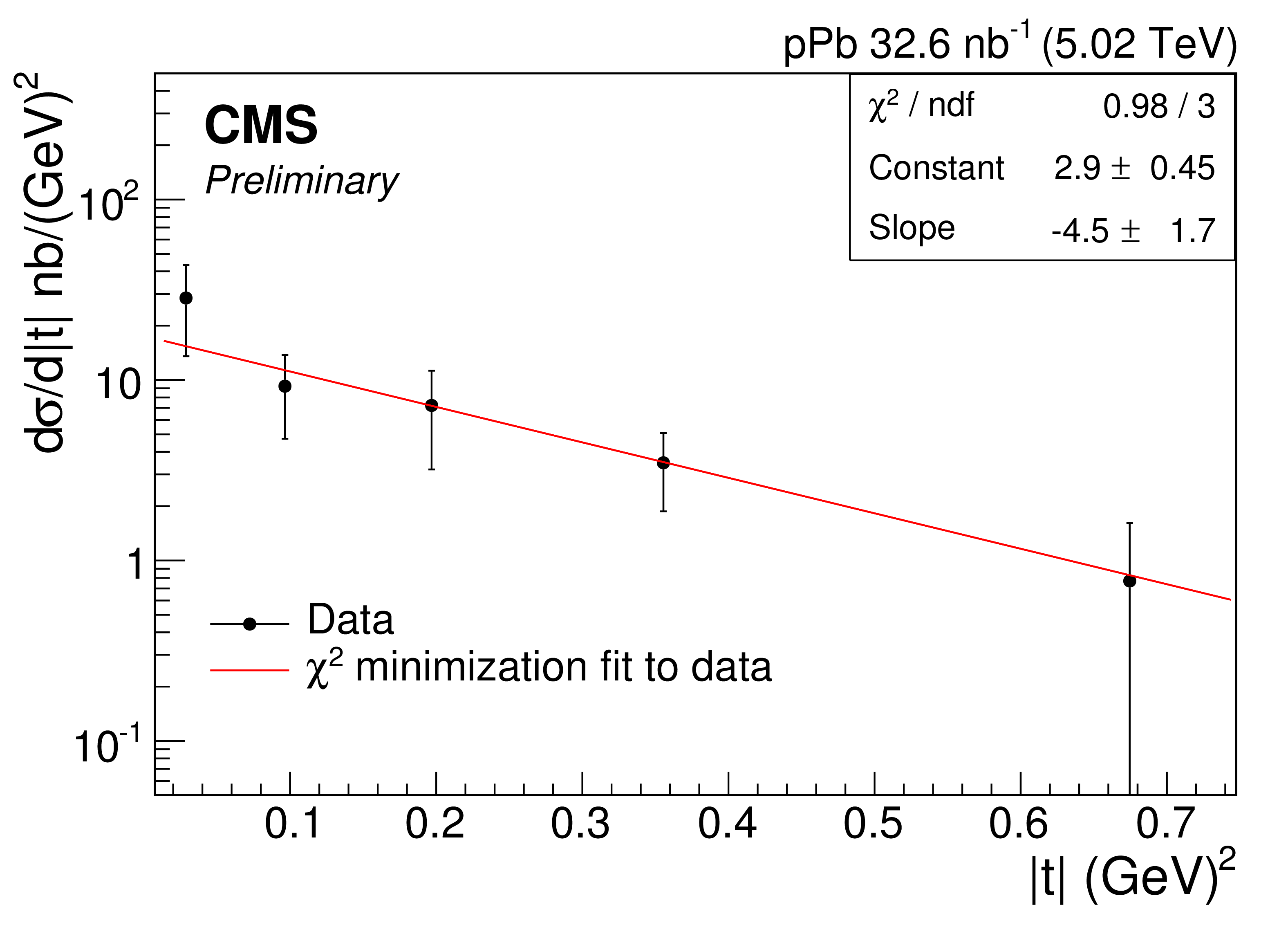}
\caption{Differential $\Upsilon$ photoproduction cross section as a function
of $|t|$ measured in pPb collisions at $\sqrt {s_{{\rm NN}}} = 5.02$ TeV in the dimuon rapidity
region $|y| < 2.2$. The solid line represents
the result of a fit with an exponential function $N e^{-b|t|}$~\cite{ref20}.}
\label{fig:pt2fit}
\end{center}
\end{figure*}

\section{Results}
\subsection{Cross section as a function of $|t|$} 
The differential cross section ${\rm d}\sigma/{\rm d}|t|$ measured 
in five bins of $|t|$ for $|y|<2.2$, as
described in section \ref{sec:xsec}, is shown in Figure~\ref{fig:pt2fit}. 
The cross section is fitted with an
exponential function $N~e^{ -b|t|}$ in the region
$0.01<|t|<1.0$ GeV$^2$, using an unbinned $\chi^2$
minimization method. A value of $b=4.5 \pm 1.7$ (stat) $\pm$ $0.6$ (syst) GeV$^{-2}$ 
is extracted from the fit. This result is in
agreement with the value $b=4.3^{+2.0}_{-1.3}$ (stat) measured by 
the ZEUS experiment~\cite{ref13} for the
photon-proton center-of-mass energy $60<W_{\gamma {\rm p}}<220$~GeV.
The measured value of $b$ is also
consistent with the predictions based on pQCD models~\cite{ref14}.  

\subsection{Cross-section as a function of $W_{\gamma p}$} 
\begin{figure*}[t]
\begin{center}
\includegraphics[width=0.80\textwidth]{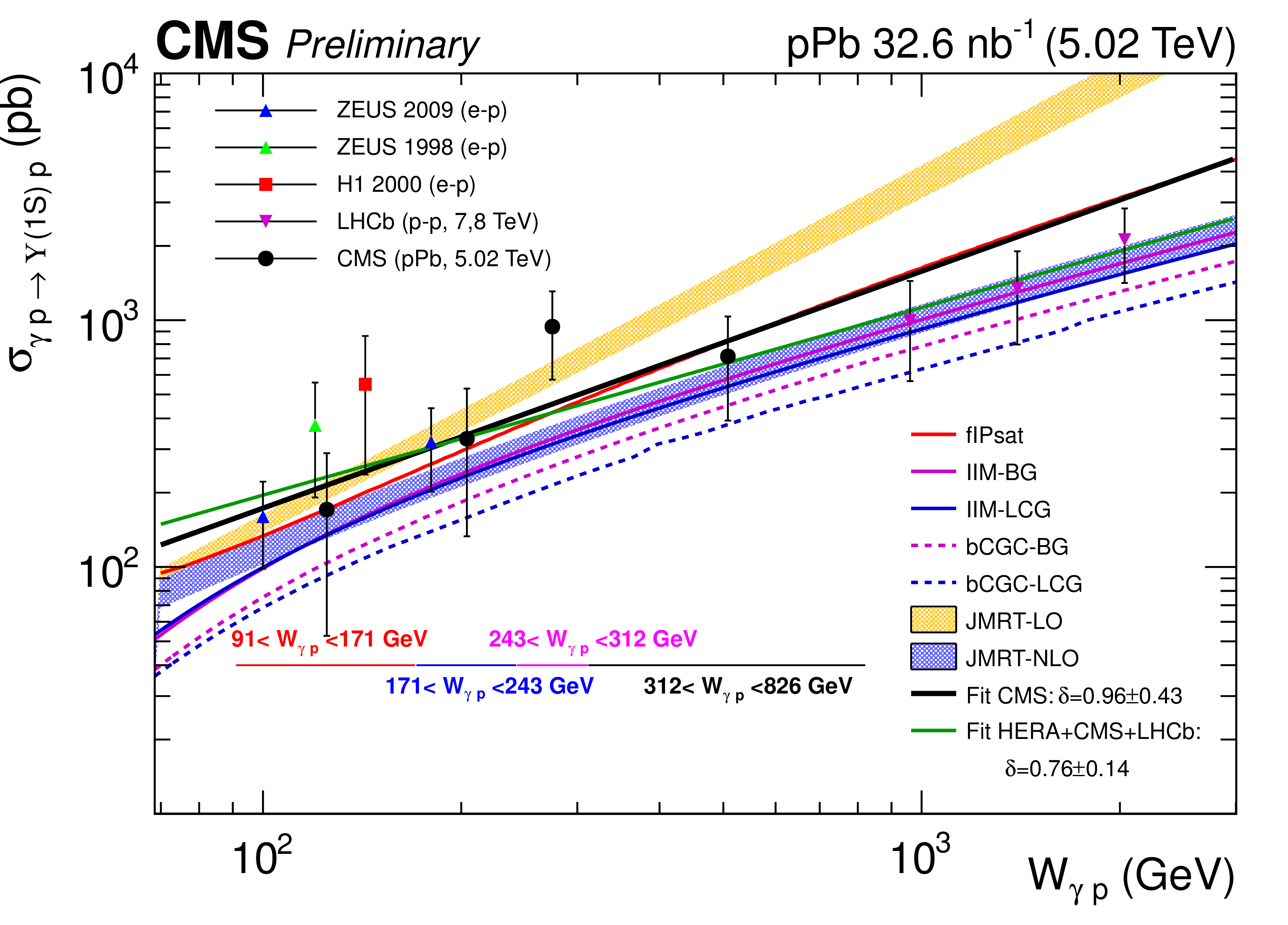}
\caption{Cross section for exclusive $\Upsilon$(1S) photoproduction, $\gamma p \rightarrow \Upsilon (1S) p$ as a function of photon-proton center-of-mass energy, $W_{\gamma p}$~\cite{ref20}.}
\label{fig:wgp}
\end{center}
\end{figure*}

The differential cross section ${\rm d}\sigma/{\rm d}y$ is measured 
in four bins rapidity, as 
described in section \ref{sec:xsec}, is shown in Figure~\ref{fig:wgp}.
The exclusive $\Upsilon$(1S)
photoproduction cross section as a function of $W_{\gamma p}$ is obtained by using, 
\begin{equation}
\sigma_{\gamma p \rightarrow \Upsilon(1S)p}(W_{\gamma p}^{2}) = \frac{1}{\Phi}\frac{d\sigma_{\Upsilon(1S)}}{dy},
\label{eq:photo_cross}
\end{equation}

where $\Phi$ is the  photon flux evaluated at the mean of the rapidity bin, 
estimated from {\textsc{Starlight}}. The CMS data are plotted together with the previous
measurements from H1~\cite{ref5}, ZEUS~\cite{ref6} and 
LHCb~\cite{ref2} data. It is also compared with different theoretical 
predictions of the JMRT model~\cite{ref14},
factorized IPsat model~\cite{ref15,ref16}, IIM~\cite{ref17,ref18} and 
bCGC model~\cite{ref19}. As $\sigma(W_{\gamma p})$ is proportional to
the square of the gluon PDF of the proton and the gluon distribution
at low Bjorken $x$ is well described by a power law,  
the cross section will also follow a power law. Any deviation 
from such trend would indicate different behaviour of gluon density function. 
We fit a power law $A\times (W/400)^\delta$ 
with CMS data alone which gives $\delta=0.96\pm 0.43$ and $A=655\pm 196$ 
and is shown by the black solid line. 
The extracted $\delta$ value is comparable to the value
$\delta=1.2 \pm 0.8$, obtained by ZEUS~\cite{ref6}.  

\section{Summary}
We reported the first measurement of the exclusive photoproduction of  $\Upsilon$(1S, 2S, 3S) mesons in the
$\mu^{+}\mu^{-}$ decay modes in ultraperipheral pPb collisions at $\sqrt {s_{{\rm NN}}} = 5.02$ TeV, using data
collected with the CMS detector in 2013, corresponding to an integrated luminosity of $33$~nb$^{-1}$.
The exclusive photoproduction cross sections have been measured as a function of the photon-proton center of mass
energy in the range $91 < W_{\gamma p}< 826$~GeV, probing the region of parton fractional momenta in the proton
$x\approx 10^{-4}$--$10^{-2}$, bridging a previously unexplored region between HERA and LHCb measurements.
Our data are compatible with a power law dependence of $\sigma(W_{\gamma p})$, disfavouring faster rising
predictions  of  LO pQCD. The differential cross section $d\sigma/d|t|$ has been also measured in the 
range $|t|<1.0$ GeV$^2$, and the exponential spectral slope $b=4.5 \pm 1.7$ (stat) $\pm$ $0.6$ (syst) GeV$^{-2}$ 
has been extracted, in agreement with earlier measurements.

\end{document}